# Bifurcated Autoencoder for Segmentation of COVID-19 Infected Regions in CT Images

Parham Yazdekhasty[1], Ali Zindar[1], Zahra Nabizadeh-ShahreBabak[1],
Roshank Roshandel[2], Pejman Khadivi[2], Nader Karimi[1], Shadrokh Samavi[1,3]

[1]Department of Elect and Computer Eng., Isfahan University of Technology, Isfahan, Iran
[2]Computer Science Department, Seattle University, Seattle, USA
[3]Department of Elect and Computer Eng., McMaster University, Hamilton, Canada

khadivip@seattleu.edu

**Abstract.** The new coronavirus infection has shocked the world since early 2020 with its aggressive outbreak. Rapid detection of the disease saves lives, and relying on medical imaging (Computed Tomography and X-ray) to detect infected lungs has shown to be effective. Deep learning and convolutional neural networks have been used for image analysis in this context. However, accurate identification of infected regions has proven challenging for two main reasons. Firstly, the characteristics of infected areas differ in different images. Secondly, insufficient training data makes it challenging to train various machine learning algorithms, including deep-learning models. This paper proposes an approach to segment lung regions infected by COVID-19 to help cardiologists diagnose the disease more accurately, faster, and more manageable. We propose a bifurcated 2-D model for two types of segmentation. This model uses a shared encoder and a bifurcated connection to two separate decoders. One decoder is for segmentation of the healthy region of the lungs, while the other is for the segmentation of the infected regions. Experiments on publically available images show that the bifurcated structure segments infected regions of the lungs better than state of the art.

## 1 Introduction

Over the past few months, COVID-19 has spread worldwide, infecting over 30 million people and causing nearly a million people loss of life [1]. The pressure on the healthcare system to both detect and treat patients has been of great concern worldwide. Medical images are used to diagnose the disease and evaluate the severity of the infection, placing a significant burden on radiologists.

Researchers have been working on machine learning approaches to classify and segment lung images to aid COVID-19 diagnosis [2]. However, the accuracy and speed of the detection process have been of great concern.

---

Parham Yazdkhasty and Ali Zindari have equal contributions

Our work has mainly focused on computed tomography (CT) images from the lungs. We will review several segmentation methods used to partition infected lung regions from healthy regions in the following. Accurate segmentation often serves as a preliminary step for further image processing and analysis [3].

One of the classical medical image segmentation approaches is Support Vector Machine (SVM), where a supervised machine learning method classifies each image pixel. In [4], each CT image lungs' nodules region is segmented by using SVM. After introducing Convolutional Neural Networks (CNNs) by [5], a new horizon opened for medical image processing, especially for medical image segmentation. A few years later, Ronneberger invented U-Net architecture [6], which has become the foundation for many advances over the last few years. The challenges with the segmentation of lung lesions in CT images predate the Coronavirus pandemic. For example, in [7], the authors combined the powerful nonlinear feature extraction ability of CNNs and the accurate boundary delineation ability of Active Contour Models (ACMs) for segmenting lesion regions in the brain, liver, and lung. Amyar et al. [8] proposed a Multi-Task Learning (MTL) approach for segmentation and COVID-19 CT images classification. Deploying MTL can solve the problem of lacking data and could lead to a model with better generalization. The authors in [8] employ a shared encoder and two parallel networks to classify infected patients and segment infected regions.

Another approach for segmenting 3D-CT images is to use a 3D U-Net [9], the same method used in [10]. It is a regular U-Net, but its 2D layers are replaced with 3D convolutional layers, which gives the model the ability to process a 3D CT image. Nevertheless, it is always a good idea to customize U-Net's base architecture for more complex tasks. In [11], Tongxue Zhou et al. proposed a U-Net based architecture, which in each step, simple 2D convolutions are followed by a "res_dil" block -- a residual architecture that is inspired from *ResNet* [12]. Also, in [11], They applied a new loss function to improve small Region of Interest (ROI) segmentation performance. They also took advantage of the *attention mechanism,* which assigns weights to emphasize important information. In their work, channel-wise and space-wise methods were used to avoid the model from distraction by non-important parts. Instead of max pooling, which is usually used for reducing the resolution of feature maps, authors in [11] have used convolution with different stride.

In some tasks, such as the lesion segmentation task, extracting textures can be helpful. In [13], this fact is used to improve the output of the segmentation task. As a preprocessing method, they segment the lungs region from CT Images. They then used similar architecture to predict the segmentation map for infected regions using the last part's output and the original image as inputs. Finding the mask for the lungs region will lead to higher segmentation accuracy.

In the early days of the COVID-19, authors of [14] focused on supporting the healthcare providers for diagnosing infections with deep learning. They used MTL to overcome challenges associated with small datasets. In this case, the primary contribution was a classification approach, but they first extracted the COVID-19 infected regions. For better prediction, the segmentation task was done by cascading two encoder-decoder based models. The first one is responsible for finding a mask for the lungs, and the second one uses the output of the previous model to segment infected regions.

In [15], each convolutional layer was replaced with a ResNet block, which is a residual block. The authors believe that the original U-Net is not "deep" enough for this problem; therefore, they made the architecture deeper by adding another deep section. These ideas improved the accuracy of prediction by about 10% compared to the original U-Net.

Laradji et al. [16] proposed a weakly-supervised method for segmentation. Using this technique reduces the cost of collecting data for training and the time that an expert should spend. Given that the flipped image segmentation map is the same as flipping the original image segmentation map, the authors of [16] use flipped images to augment their training dataset. However, we believe that the model used in [16] could perform better if the lungs segmentation-maps were used.

In this paper, we propose a new bifurcated architecture for segmenting infected regions of COVID-19. Our architecture consists of an encoder and two decoders. Our approach relies on MTL because we have a small dataset of images for this task. Our proposed architecture learns how to segment the lungs and infected regions simultaneously.

The structure of this paper is as follows. In Section 2, the proposed method is explained. The metrics that are used for evaluation are described in Section 3. Then in Section 4, the experimental results are reported. In the final section, we conclude the article.

## 2      Proposed Method

Segmenting infected regions in CT images can help speed up the detection of COVID-19 infections. Due to the different sizes of infected regions and small datasets, achieving accurate results is difficult. In this paper, a new architecture for segmenting COVID-19 infected regions is proposed. This architecture is based on U-Net [6], which is previously used for segmenting medical images. It consists of four blocks: a *decoder*, a bifurcated structure with *two parallel encoders*, and a final *merging encoder*. The block diagram of our models is shown in Fig. 1. In the following, the details of the proposed architecture are explained.

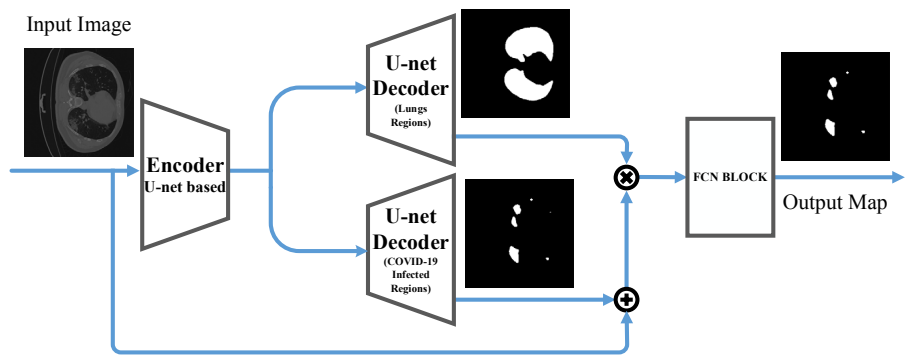

**Fig. 1**. The proposed bifurcated architecture.

## 2.1 Architecture

One technique that we consider in our architecture is Multi-task learning (MTL) [17]. This technique is a learning algorithm to train a network for multiple related tasks. The basic idea is that multiple related tasks can have shared features. In such a case, the model learns which features are common and which features are distinct for each class. Hence, the problem of overfitting is avoided.

In MTL, deep learning is typically performed using either hard parameter sharing or soft parameter sharing. In hard parameter sharing, hidden layers are shared among all tasks while keeping several task-specific output layers. On the other hand, in soft parameter sharing, each task has its network architecture (model and parameters), and there is no shared layer between these tasks. In soft sharing, the distance between network parameters is regularized to make similar parameters for different tasks.

In this paper, we use MTL to increase the segmentation accuracy of infected regions. However, contrary to other works, both tasks in our work are segmentation, one for the lungs region and one for infected regions. The encoder is shared for both tasks, but each task has a separate decoder. Since our architecture's encoder is common for the bifurcated branches, parameters are shared, and the complexity is reduced. In the following, the structure of each block that is shown in Fig. 1 is explained.

### 2.1.1. The encoder

As a regular U-Net architecture, our network begins with an encoder responsible for extracting useful feature maps to be used by the decoder later on. Our approach extracts both high-level and low-level features. We use a modified inception block [18] in the encoder's front layers when the image resolution is still high. As the feature maps go through the pooling layers, their sizes and resolutions are reduced. An inception block is shown in Fig. 2.

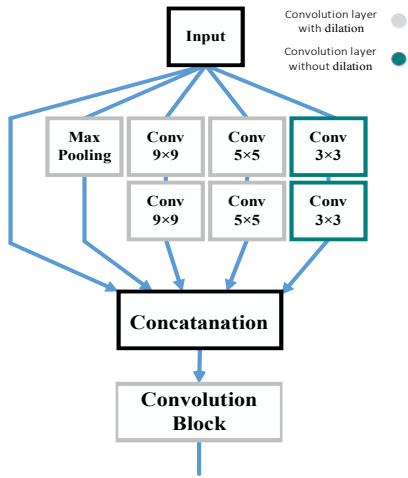

**Fig. 2**. Proposed modified inception block.

The inception block applies filters with different size kernels to the same input, rather than applying a stack of same-size kernel convolutional layers. The use of filters with multiple sizes enables us to extract different feature maps with various scales. Intuitively, the infected regions' sizes are different, so these feature maps help the model have better predictions by having different receptive fields.

There are four branches in our inception block:
   a) Two 3×3 convolutional layers without dilation
   b) Two 5×5 convolutional layers with dilation rate of "*d_rate*"
   c) Two 9×9 convolution layers with dilation rate of "*d_rate*"
   d) Max-pooling 2D with a size of 7×7 and a stride of 1×1.

Given that there are three inception blocks in our proposed architecture. In this inception block, two convolutional blocks are dilated type. The advantage of this type of convolution is having an extended field of view while the computational cost is not increased. For this purpose, a parameter determines the distance of neighbors from the center of the kernel. In our work, this parameter is called "d_rate." We determine the value for "*d_rate*" based on the corresponding inception block position in the network. For the inception blocks located toward each decoder's end, smaller "*d_rate*" values are used to allow for more conservative predictions.

### 2.1.2. Segmentation Decoders

The bifurcated part of the architecture is responsible for constructing desirable outputs, which have the same size as the input image. The decoder on the upper branch (Fig. 1) is responsible for the lung region's segmentation. The output of this upper decoder is a binary mask that specifies where the lungs are located. We will use the binary lungs mask as a region of interest (ROI) in our architecture's next stage. The decoder in the lower bifurcated branch is used to segment the infected regions of the lungs. This decoder's primary duty is to generate a probability map for the next stage to extract the COVID-19 infected regions.

Another reason for using the bifurcation structure is implementing the *deep supervision* technique. The deep supervision method is proposed in [19] for preventing the vanishing gradient descent. In addition to the loss generated at the output, deep supervision uses auxiliary loss from different network layers. We deployed the deep supervision method to train the network to predict infected regions. In our architecture, before the final output, the auxiliary loss function is defined at the end of the lower decoder. At this point, we have two loss functions in this block that help the model to learn features and avoid vanishing gradient. The loss function and the ground truth of the lower decoder are the same as the final output.

The architectures of the two decoders are very similar. Each decoder is a U-Net decoder that uses features extracted by the encoder. Skip connections are not shown between layers of the encoder and decoders. At the end of these decoders, another inception block is employed to let the decoder make its final decisions using multi-scale kernels in convolutions.

### 2.1.3. Merging outputs and FCN block

The Fully Convolutional Networks (FCN) block is the last part of our architecture that produces the final output. The FCN gets its inputs from the two segmentation decoders. To achieve accurate segmentation results, we use probability maps before binarization instead of using the infected regions' binary segmentation map. We then concatenate the input image and the probability map. The lung's binary output from the upper decoder is multiplied by the above-mentioned concatenated outputs for removing the misleading information. The output of multiplication is then fed into an FCN layer.

### 2.2 Loss Functions

Our loss function consists of three parts. The first loss function is from the lung decoder. The second loss function is from the infected region decoder. The third loss function is for the FCN block. We found that binary cross-entropy is the best loss function for classification and segmentation tasks. The formula of binary cross-entropy is as follow:

$$L = \frac{-1}{N}\sum_{i=1}^{N} y_i log(p(y_i)) + (1 - y_i).log(1 - p(y_i))$$

where $N$ is the number of pixels, $y_i$ is the label of the $i$-th pixel, and $p(y_i)$ is the predicted label.

Experimental results show that the effect of each of the three parts is different. Therefore, we use a weighted binary cross-entropy loss function. We use a weight of 0.5 for the lungs segmentation ($W_{lung}$), 1 for the auxiliary output ($W_{aux}$), and 2 for the final output ($W_{fin}$).

$$Loss = W_{lung}L_{lung} + W_{aux}L_{aux} + W_{fin}L_{fin}$$

## 3 Experimental Results

To evaluate our architecture and compare it with other state-of-the-art algorithms, we have used various standard metrics shown in Table 1.

**Table 1.** The metrics for evaluating

| Metric | Formula |
|---|---|
| Sensitivity | $\dfrac{TP}{TP + FN}$ |
| Specificity | $\dfrac{TN}{TN + FP}$ |
| IoU | $\dfrac{TP}{TP + FP + FN}$ |
| Dice Score | $\dfrac{2*TP}{2*TP + FP + FN}$ |
| PPV | $\dfrac{TP}{TP + FP}$ |

We have used the standard definition for these metrics based on true/false negative/positive rates as in Table 1. Specifically, we use *sensitivity* because we are interested in evaluating how robust and trustworthy is our model in detecting infected regions. Intersection over Union (IoU) and Dice Score are commonly used in evaluating segmentation tasks to calculate the overlap between the ground truth and the region predicted by the algorithm. For emphasizing the positive samples, we used PPV (Positive Predicted Value).

We use a publicly available dataset [20] used for similar tasks by [21] to evaluate our work. This dataset contains 20 labeled COVID-19 CT scans. Left lung, right lung, and infections are labeled by a radiologist and verified by an experienced radiologist. Each instance of the dataset is a 3D-volume consisting of multiple slices. Since our dataset's images have different sizes, we resized each slice to 256×256 pixels to train the model. Furthermore, to address computation limitations, the 3D dataset was converted into 2D slices. The 3D CT images in the dataset have different slices and require a network capable of accepting 3D tensors. By using the 2D dataset, a simpler network can be used. The 2D dataset contains 3520 slices. About 52% (1841) of slices correspond to infected regions of the lungs, while the rest are images of healthy regions. We then selected 16 random training sets and four test sets. We selected 12% of training slices randomly as our validation set due to the slice level processing. We experimented with a five-fold and ran the model three times on each fold. The average number of slices in different folds for the training set is 2816, and the test set is 704. The average result is reported in this section.

As mentioned earlier, the last layer of our model's output is the class probability value for each pixel. A threshold should be selected to convert the probability into a binary value and generate the segmented mask. We use our validation data and experiment with different thresholds to obtain a suitable marker for selecting this threshold. We then apply this threshold to our test data. To evaluate our model's performance, we compare our results with [16], which uses weakly supervised learning, and [22], which uses a 3D U-Net model for solving this problem.

Table 2 shows the results of various metrics on five folds. The result of each fold is computed as the average of three runs. The last column reports the average and standard deviation of 5 the folds.

**Table 2**. Average metrics on five folds. Each fold is trained three times.

|  | Fold 1 | Fold 2 | Fold 3 | Fold 4 | Fold 5 | Average$\pm std$ |
|---|---|---|---|---|---|---|
| **IOU** | 0.719 | 0.746 | 0.594 | 0.699 | 0.752 | 0.702$\pm$0.057 |
| **Dice** | 0.778 | 0.812 | 0.669 | 0.762 | 0.819 | 0.768$\pm$0.053 |
| **PPV** | 0.831 | 0.827 | 0.764 | 0.826 | 0.852 | 0.82$\pm$0.029 |
| **Sensitivity** | 0.779 | 0.821 | 0.633 | 0.755 | 0.81 | 0.759$\pm$0.067 |
| **Specificity** | 0.998 | 0.997 | 0.997 | 0.999 | 0.998 | 0.99$\pm$0.0007 |

Table 3 compares our results with those of [22]. In [22], the authors also divided the dataset into five folds. The results of each fold for our model and the model of [22] are shown in Table 3. Our model performs better on Dice score and sensitivity metrics while slightly lagging in specificity. The dice and sensitivity outperform the [22] due to different view fields using the inception block, which helps the model detect infection regions with different sizes accurately.

**Table 3.** Comparing our method with [22]

|      | Folds   | Dice  | Sensitivity | Specificity |
|------|---------|-------|-------------|-------------|
|      | Fold 1  | 0.556 | 0.447       | 0.999       |
|      | Fold 2  | 0.801 | 0.875       | 0.999       |
| [22] | Fold 3  | 0.829 | 0.796       | 0.999       |
|      | Fold 4  | 0.853 | 0.836       | 0.999       |
|      | Fold 5  | 0.765 | 0.697       | 0.999       |
|      | Average | 0.761 | 0.730       | **0.999**   |
|      | Fold 1  | 0.778 | 0.779       | 0.998       |
|      | Fold 2  | 0.812 | 0.821       | 0.997       |
| Ours | Fold 3  | 0.669 | 0.633       | 0.997       |
|      | Fold 4  | 0.762 | 0.755       | 0.999       |
|      | Fold 5  | 0.819 | 0.81        | 0.998       |
|      | Average | **0.768** | **0.759** | 0.997     |

Table 4 compares our results with those of [16]. Except for the sensitivity, our results outperform the model in [16].

**Table 4.** Comparing our method with [16]

|      | Dice   | Sensitivity | Specificity | IOU   | PPV   |
|------|--------|-------------|-------------|-------|-------|
| [16] | 0.75   | **0.86**    | 0.97        | 0.59  | 0.66  |
| Ours | **0.768** | 0.759    | **0.997**   | **0.702** | **0.82** |

Our proposed bifurcated method allows the model to concentrate on using features specifically located in the lungs region. Otherwise, the network could have been misled by features outside the lungs. Also, using the inception block made it possible to have different scales of features, which helped achieve accurate segmentation of infected regions with different sizes. In Fig. 3, some samples of the dataset and the results of our models are shown. In this figure, the first column is the input image, the second column is the segmented results of our model, the third column is the ground truth, and the last one is the combined input image with segmented results. Since the infected regions have different sizes, our model could accurately segment the large regions, but some small regions are missed.

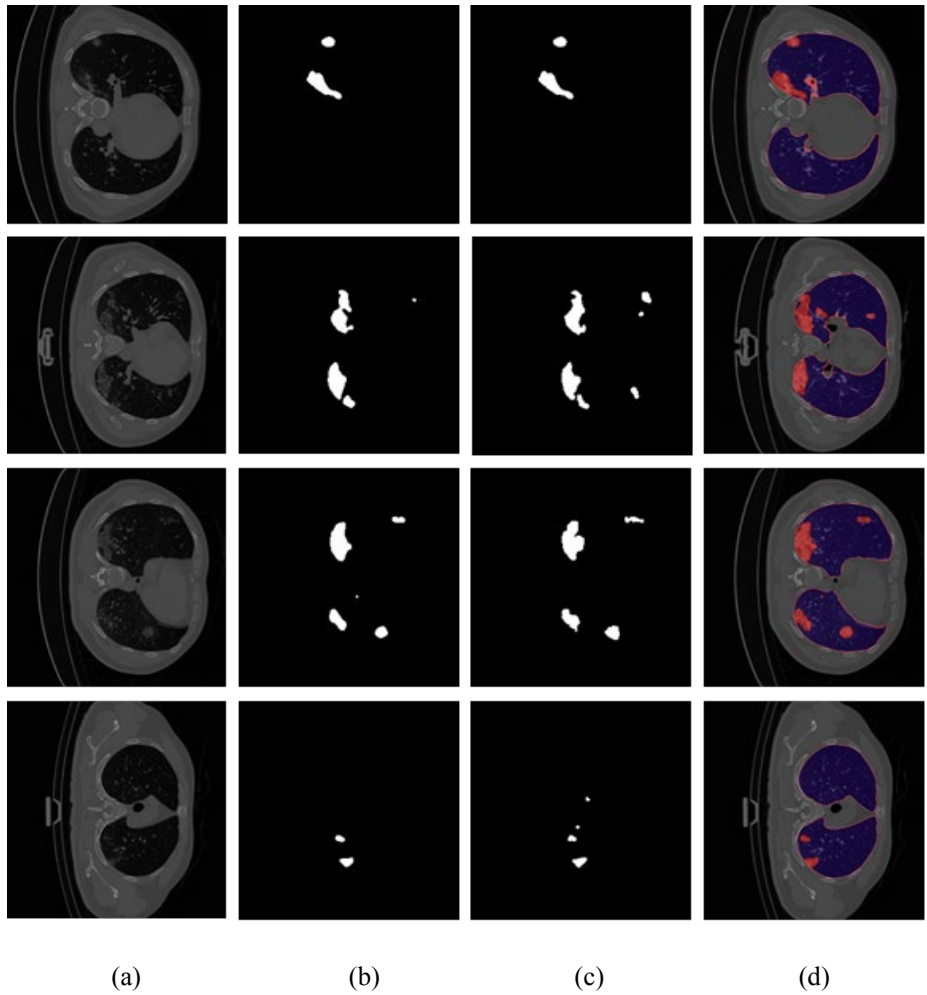

(a)          (b)          (c)          (d)

**Fig. 3.** (a) input images, (b) predicted mask, (c) ground truth, (d) predicted mask superimposed on the input image.

## 4      Conclusion

In this paper, we proposed a model for segmenting COVID-19 infected lung regions in CT imaging. In our method, an encoder is shared by two decoders. The output of the encoder is bifurcated and is fed two separate decoders. The two decoders' output is combined and used to predict the infected lungs areas' final segmentation map.

For more accurate and precise predictions, the model extracts the lungs' mask of the given input to avoid the image's useless regions. By using the inception blocks, the infected regions are segmented more accurately. Also, using deep supervision helped the model to learn better. The proposed method could be categorized as a multi-task learning system. By using these techniques in our model, the results surpassed the state of the art.